\documentclass[aps,twocolumn,pra,showpacs,floatfix]{revtex4}

\usepackage[dvips]{graphicx}

\begin{document}


\title{Frequency shift of cesium clock transition due to blackbody radiation}

\author{E. J. Angstmann}
\affiliation{School of Physics, University of New South Wales,
Sydney 2052, Australia}
\author{V. A. Dzuba}
\affiliation{School of Physics, University of New South Wales,
Sydney 2052, Australia}
\author{V. V. Flambaum}
\affiliation{School of Physics, University of New South Wales,
Sydney 2052, Australia}

\date{\today}

\begin{abstract}
We have performed {\em ab initio} calculations of the frequency
shift induced by a static electric field on the cesium clock
hyperfine transition. The calculations are used to find the
frequency shifts due to blackbody radiation. Our result ($\delta
\nu/E^2=-2.26(2)\times 10^{-10}$Hz/(V/m)$^2$) is in good agreement
with early measurements and {\em ab initio} calculations performed
in other groups. We present arguments against recent claims that the
actual value of the effect might be smaller. The difference ($\sim$
10\%) between {\em ab initio} and semiempirical calculations is due
to the contribution of the continuum spectrum to the sum over
intermediate states.
\end{abstract}

 \maketitle

Atomic clocks are now important for both practical applications and fundamental
physics. One of the dominant uncertainties in high-precision measurements
of frequencies in atomic clocks is the ac stark
shift induced by blackbody radiation (see e.g. \cite{Itano}).
There is some disagreement on the value of this shift.
Early measurements~\cite{Haun,Mowat,Simon} and {\em ab initio}
calculations~\cite{Lee,Palchikov} support a value which is close to
$-2.2\times 10^{-10}$Hz/(V/m)$^2$ while more recent
measurements~\cite{Levi,Godone} and semiempirical
calculations~\cite{Feitchner,Micalizio,Ulzega} claim that actual
number might be about 10\% smaller.

In the present work we have performed fully {\em ab initio}
calculations of the radiation frequency shift and have identified
the source of the disagreement between different theoretical results
as the contribution of the continuum spectrum states into summation
over the complete set of intermediate states. The continuum spectrum
was included in all the {\em ab initio} calculations and missed in
the semiempirical considerations. We demonstrate that adding the
contribution of the continuum spectrum to where it was missed brings
all theoretical results to good agreement with each other and with
early measurements.

Blackbody radiation creates a temperature dependent electric field,
described by the Planck radiation law
\begin{equation}\label{Planck}
E^{2}(\omega) = \frac{8 \alpha}{\pi}\frac{\omega ^{3} d
\omega}{\textrm{exp}(\omega/kT)-1}.
\end{equation}
This leads to the following expression for the average electric
field radiated by a black body at temperature T:
\begin{equation}\label{electric_field}
    \langle E^{2} \rangle = (831.9 \textrm{V/m})^{2}[\textrm{T(K)/300}]^{4}.
\end{equation}
This electric field causes a temperature-dependent frequency shift
of the atomic microwave clock transitions. It can be presented in
the form (see, e.g.~\cite{Itano})
\begin{equation}\label{beta}
    \delta \nu/\nu_0 = \beta (T/T_0)^4 \left[ 1 + \epsilon (T/T_0)^2 \right]
\end{equation}
Here $T_0$ is usually assumed to be room temperature ($T_0 = 300K$).
The frequency shift in a static electric field is
\begin{equation}\label{ke2}
    \delta \nu = k E^2.
\end{equation}
Coefficients $k$ and $\beta$ are related by
\begin{eqnarray}
    \beta &=& \frac{k}{\nu_0} (831.9 \textrm{V/m})^2 \\
          &=& k \times7.529\times 10^{-5} (\textrm{V/m})^2\textrm{Hz}^{-1} \ \ (\textrm{for~Cs}), \nonumber
\end{eqnarray}
while $\epsilon$ is a small correction due to frequency distribution
(\ref{Planck}). In the present work we calculate the coefficient
$k$.

In the case when there is no other external electric field the
radiation shift can be expressed in terms of the scalar hyperfine
polarizability of the atom. This corresponds to averaging over all
possible directions of the electric field. The hyperfine
polarizability is the difference in the atomic polarizabilities
between different hyperfine structure states of the atom. The
lowest-order effect is linear in the hyperfine interaction and
quadratic in the electric field. The corresponding third-order
perturbation theory expressions, after angular reduction have the
form
\begin{eqnarray}
\label{nu1}
 && \delta \nu_1(as) = e^2 \langle E^2 \rangle \frac{2I+1}{6} \times \nonumber \\
 && \sum_{n,m,j}
   \frac{ A_{as,ns} \langle ns || r || mp_j \rangle \langle mp_j || r || as \rangle}
  {(\epsilon_{as} - \epsilon_{ns})(\epsilon_{as} - \epsilon_{mp_j})},
\end{eqnarray}

\begin{eqnarray}
\label{nu2}
 && \delta \nu_2(as) = \frac{e^2 \langle E^2 \rangle }{6}
  \sum_{j}(C_{I+1/2} - C_{I-1/2}) \times \nonumber \\
 && \sum_{n,m} \frac{\langle as||r||npj \rangle
   A_{npj,mpj}  \langle mp_j || r || as \rangle}
  {(\epsilon_{as} - \epsilon_{npj})(\epsilon_{as} - \epsilon_{mp_j})},
\end{eqnarray}
and
\begin{eqnarray}
\label{norm}
 &&\delta \nu_{norm}(as) = \\
 &&- e^2 \langle E^2 \rangle \frac{2I+1}{12}
  A_{as} \sum_{m,j}
    \frac{|\langle as || r || mp_j \rangle|^2}
  {(\epsilon_{as} - \epsilon_{mp_j})^2}. \nonumber
\end{eqnarray}
Here
\begin{eqnarray}
&&C_F = \sum_{F'} (2F'+1)[F'(F'+1)-I(I+1)-j(j+1)] \nonumber \\
&& \times \left\{ \begin{array}{ccc} 1/2 & F & I \\ F' & j & 1
\end{array} \right\}^2, \ \ \ F'=|I-J|, I+J,
\nonumber
\end{eqnarray}
$A_{ns}$ is the hfs constant of the $ns$ state, $A_{m,n}$ is the
off-diagonal hfs matrix element, $I$ is nuclear spin,
$\mathbf{F=I+J}$, $\mathbf{J}$ is total electron momentum of the
atom in the ground state ($J=1/2$), and $j$ is total momentum of virtual $p$-states
($j=1/2,3/2$). Summation goes over a complete set of $ns$,
$mp_{1/2}$ and $mp_{3/2}$ states.

In order to calculate frequency shift to the hfs transitions due to
the electric field one needs to have a complete set of states and to
have the energies, electric dipole transition amplitudes and
hyperfine structure matrix elements corresponding to these states.
It is possible to consider summation over the physical states and to
then use experimental data to perform the calculations. The lowest
valence states for which experimental data is usually available
dominate in the summation. Off-diagonal hfs matrix elements can be
obtained to a high accuracy as the square root of the product of
corresponding hfs constants: $A_{m,n} = \sqrt{A_m A_n}$ (see,
e.g.~\cite{offdhfs}). However, the accuracy of this approach is
limited by the need to include the {\it tail} contribution from
highly excited states including states in the continuum. This
contribution can be very significant and its calculation is not
easier than the calculation of the whole sum.

Therefore, in the present work we use an {\em ab initio} approach in
which high accuracy is achieved by the inclusion of all important
many-body and relativistic effects. We make only one exception
toward the semiempirical approach. The frequency shift is dominated
by the term (\ref{norm}) which is proportional to the hfs in the
ground state.  It is natural to use experimental hfs constant in the dominating term
to have more accurate results. Note however that the difference with
complete {\it ab initio} calculations is small.

Calculations start from the relativistic Hartree-Fock (RHF) method
in the $V^{N-1}$ approximation. This means that the initial RHF
procedure is done for a closed-shell atomic core with the valence
electron removed. After that, states of the external electron are
calculated in the field of the frozen core. Correlations are
included by means of the correlation potential method~\cite{CPM}. We
use two different approximations for the correlation potential,
$\hat \Sigma$. First, we calculate it in the lowest, second-order of
the many-body perturbation theory (MBPT). We use notation $\hat
\Sigma^{(2)}$ for the corresponding correlation potential. Then we
also include into $\hat \Sigma$ two classes of the higher-order
terms: screening of the Coulomb interaction and hole-particle
interaction (see, e.g.~\cite{all-order} for details). These two
effects are included in all orders of the MBPT and the corresponding
correlation potential is named $\hat \Sigma^{(\infty)}$.

To calculate $\hat \Sigma^{(2)}$ we need a complete set of
single-electron orbitals. We use the B-spline technique
\cite{Johnson1, Johnson3} to construct the basis. The orbitals are
built as linear combinations of 50 B-splines in a cavity of radius
40$a_B$. The coefficients are chosen from the condition that the
orbitals are eigenstates of the RHF Hamiltonian $\hat H_0$ of the
closed-shell core. The $\hat \Sigma^{(\infty)}$ operator is
calculated with the technique which combines solving equations for
the Green functions (for the direct diagram) with the summation over
complete set of states (exchange diagram)~\cite{all-order}.

The correlation potential $\hat \Sigma$ is then used to build a new
set of single-electron states, the so-called Brueckner orbitals.
This set is to be used in the summation in equations (\ref{nu1}),
(\ref{nu2}) and (\ref{norm}). Here again we use the B-spline
technique to build the basis. The procedure is very similar to the
construction of the RHF B-spline basis. The only difference is that
new orbitals are now the eigenstates of the $\hat H_0 + \hat \Sigma$
Hamiltonian, where $\hat \Sigma$ is either $\hat \Sigma^{(2)}$ or
$\hat \Sigma^{(\infty)}$.

Brueckner orbitals which correspond to the lowest valence states are
good approximations to the real physical states. Their quality can
be checked by comparing experimental and theoretical energies.
Moreover, their quality can be further improved by rescaling the
correlation potential $\hat \Sigma$ to fit experimental energies
exactly. We do this by replacing the $\hat H_0 + \hat \Sigma$ with
the $\hat H_0 + \lambda \hat \Sigma$ Hamiltonian in which the
rescaling parameter $\lambda$ is chosen for each partial wave to fit
the energy of the first valence state. The values of $\lambda$ are
$\lambda_s =0.8$ and $\lambda_p =0.85$ for $\hat \Sigma^{(2)}$ and
$\lambda_s =0.99$ and $\lambda_p =0.95$ for $\hat \Sigma^{(\infty)}$.
Note that the values are very close to
unity. This means that even without rescaling the accuracy is very
good and only a small adjustment to the value of $\hat \Sigma$ is
needed. Note also that since the rescaling procedure affects both
energies and wave functions, it usually leads to improved values of
the matrix elements of external fields. In fact, this is a
semiempirical method to include omitted higher-order correlation
corrections.

Matrix elements of the hfs and electric dipole operators are found by means of the
time-dependent Hartree-Fock (TDHF) method~\cite{CPM,TDHF}. This method is equivalent
to the well-known random-phase approximation (RPA). In the TDHF method, single-electron
wave functions are presented in the form $\psi = \psi_0 + \delta \psi$, where $\psi_0$
is unperturbed wave function. It is an eigenstate of the RHF Hamiltonian $\hat H_0$:
$(\hat H_0 -\epsilon_0)\psi_0 = 0$.  $\delta \psi$ is the correction due to external
field. It can be found be solving the TDHF equation
\begin{equation}
    (\hat H_0 -\epsilon_0)\delta \psi = -\delta\epsilon \psi_0 - \hat F \psi_0 - \delta \hat V^{N-1} \psi_0,
    \label{TDHF}
\end{equation}
where $\delta\epsilon$ is the correction to the energy due to external field ($\delta\epsilon\equiv 0$
for the electric dipole operator), $\hat F$ is the operator of the external field
($\hat H_{hfs}$ or $e\mbox{\boldmath{$E$}}\cdot \mbox{\boldmath{$r$}}$), and $\delta \hat V^{N-1}$ is
the correction to the self-consistent potential of the core due to external field.
The TDHF equations are solved self-consistently for all states in the core. Then
matrix elements between any (core or valence) states $n$ and $m$ are given by
\begin{equation}
    \langle \psi_n | \hat F + \delta \hat V^{N-1} | \psi_m \rangle.
    \label{mel}
\end{equation}
The best results are achieved when $\psi_n$ and $\psi_m$ are Brueckner orbitals
calculated with rescaled correlation potential $\hat \Sigma$.

We use equation (\ref{mel}) for all hfs and electric dipole matrix
elements in (\ref{nu1}), (\ref{nu2}), and (\ref{norm}) except for
the ground state hfs matrix element in (\ref{norm}) where we use experimental data.

To check the accuracy of the calculations we perform calculations of
the hfs in the ground state and of the static scalar
polarizability. Polarizability is given by the expression
\begin{equation}
    \alpha_0(a) =\frac{2}{3} \sum_m \frac{|\langle a || r || m \rangle|^2}
  {\epsilon_{a} - \epsilon_{m}}
  \label{alpha0}
\end{equation}
which is very similar to the term (\ref{norm}) for the frequency
shift. The most important difference is that the energy denominator
is squared in term (\ref{norm}) but not in (\ref{alpha0}). This
means better convergence with respect to the summation over complete
set of states for term (\ref{norm}) than for (\ref{alpha0}).
Therefore, if good accuracy is achieved for polarizabilities, even
better accuracy should be expected for the term (\ref{norm}) (see
also Ref.~\cite{Micalizio}).

However, the behavior of the other two terms, (\ref{nu1}) and
(\ref{nu2}), is very different and calculation of polarizabilities
tells us little about accuracy for these terms.
Therefore, we also perform detailed calculations of the hfs
constants of the ground state. Inclusion of core polarization
(second term in (\ref{mel})) involves summation over the complete
set of states similar to what is needed for term (\ref{nu1}).
Comparing experimental and theoretical hfs is a good test for the
accuracy for this term.

In addition to term (\ref{mel}), we also include two smaller
contributions to the hfs: structure radiation and the correction due
to the change of the normalization of the wave function.
Our final result for the hfs constant is 2278~MHz which is in excellent agreement with the experimental value 2298~MHz~\cite{Fuller}. The result for static polarizability is $\alpha_0 = 399.0~a_0^3$ which is also in a very good agreement with experimental value $401.0(6)~a_0^3$~\cite{Amini}.

Table \ref{terms} presents contributions of terms (\ref{nu1}),
(\ref{nu2}) and (\ref{norm}) into the total frequency shift of the
hfs transition for the ground state of $^{133}$Cs calculated
in different approximations. Term (\ref{norm}) dominates while term
(\ref{nu2}) is small but still important. Results obtained with $\hat \Sigma^{(2)}$ and
$\hat \Sigma^{(\infty)}$ differ significantly (14\%).
However, after rescaling the results for both $\hat \Sigma^{(2)}$
and $\hat \Sigma^{(\infty)}$ come within a fraction of a per cent of
each other. Naturally, rescaling has a larger effect on results
obtained with $\hat \Sigma^{(2)}$. This means that the rescaling
really imitates the effect of higher-order correlations and should
lead to more accurate results.

In summary, we have three ways of estimation of the accuracy of calculations: (a) calculation of static polarizability (0.5\% accuracy); (b) calculation of the hfs (0.9\% accuracy); and (c) comparision of the results obtained in different most accurate approximations (three last lines of Table~\ref{terms}), which differ by about 0.3\%. Therefore, we can say that the accuracy of the calculations is not worse that 1\%.
Our final result is
\begin{equation} \label{final}
    k = -2.26(2) \times 10^{-10} \textrm{Hz} /(\textrm{V/m})^2.
\end{equation}
This corresponds to $\beta = -1.70(2) \times 10^{-14}$. To obtain frequency shift at finite temperature one needs to substitute this value into equation~(\ref{beta}). For accurate results one also needs to know the value of $\epsilon$. It was estimated in Ref.~\cite{Itano} in single-resonance approximation and found to be 0.014. In many-resonance calculation $\epsilon$ will be 10-20\% smaller.

\begin{table}
    \caption{Contribution of terms (\ref{nu1}), (\ref{nu2}), and (\ref{norm}) to the frequencies of the hyperfine
    transition in the ground state of  $^{133}$Cs ($\delta\nu_{0}$/$E^{2} \times 10^{-10}$ Hz/(V/m)$^2$) in different approximations.}
    \begin{ruledtabular}
        \begin{tabular}{l r r r r r}
     \multicolumn{1}{c}{$\hat \Sigma$} & \multicolumn{1}{c}{(\ref{nu1})} & \multicolumn{1}{c}{(\ref{nu2})} &
        \multicolumn{1}{c}{(\ref{norm})} & \multicolumn{1}{c}{Total}  \\
        \hline

         $\hat \Sigma^{(2)}$\footnotemark[1]              & -0.9419 & 0.0210 & -1.0722 & -1.9931 \\
         $\lambda \hat \Sigma^{(2)}$\footnotemark[2]      & -1.0239 & 0.0229 & -1.2688 & -2.2697 \\
         $\hat \Sigma^{(\infty)}$\footnotemark[3]         & -1.0148 & 0.0232 & -1.2706 & -2.2622 \\
         $\lambda \hat \Sigma^{(\infty)}$\footnotemark[2] & -1.0167 & 0.0230 & -1.2695 & -2.2632 \\

       \end{tabular}
    \footnotetext[1]{$\hat \Sigma^{(2)}$ is the second-order correlation potential.}
    \footnotetext[2]{Rescaled $\hat \Sigma$.}
    \footnotetext[3]{$\hat \Sigma^{(\infty)}$ is the all-order correlation potential.}
      \label{terms}
        \end{ruledtabular}
\end{table}

We present our final result for the frequency shift together with other theoretical and experimental results in Table~\ref{results}. Our value is in good
agreement with early measurements~\cite{Haun,Mowat,Simon} and {\em
ab initio} calculations~\cite{Lee,Palchikov} while recent
measurements~\cite{Levi,Godone} and semiempirical
calculations~\cite{Feitchner,Micalizio,Ulzega} give the value which
is about 10\% smaller. Less accurate measurements of Bauch and
Schr\"{o}der~\cite{Bauch} cover both cases. We cannot comment on
disagreement between experimental results. However, the source of
disagreement between theoretical results seems to be clear. It comes
from the contribution of the continuum spectrum to the summation
over the complete set of states in term (\ref{nu1}). This term has
off-diagonal hfs matrix elements between the ground state and
excited states. Since the hfs interaction is localized over short
distances ($ \sim a_0/Z$) it emphasizes the contribution of states
with high energies including states in the continuum (since $\Delta
p \Delta x \geq\hbar$, a small area of localization ($\Delta x$)
allows high momentum ($p$) and thus high energy). In our
calculations the contribution of states above $7p$ into term
(\ref{nu1}) is $-0.35 \times 10^{-1}$Hz/(V/m)$^2$ which is 15\% of
the total answer.

In contrast, states above $7p$ contribute only about 0.05\% of the
total value of term (\ref{norm}). This is because the summation goes
over the matrix elements of the electric dipole operator which is
large on large distances and thus suppresses the contribution of
high-energy states. It is not surprising therefore that a
semiempirical consideration, involving only discrete spectrum
states, gives very good results for the atomic polarizabilities
(see, e.g.~\cite{Micalizio}). However, let us stress once more that
the calculation of polarizabilities checks only term (\ref{norm})
and tells us very little about the accuracy of other two terms,
(\ref{nu1}) and (\ref{nu2}).

The contribution of the states above $7p$ is even more important for
term (\ref{nu2}). Their contribution is about 30\% of the total
value of this term. However, the term itself is small and its
accurate treatment is less important.

In {\it ab initio} calculations by Lee {\it et al}~\cite{Lee}
summation over complete set of states is reduced to solving of a
radial equation (equations of this type are often called Sternheimer
equation after one of the authors of this work). This approach does
include the contribution of the continuum spectrum and the result is
in very good agreement with ours (see Table~\ref{results}).

In other {\it ab initio} calculations by Pal'chikov {\it et
al}~\cite{Palchikov} summation is done via Green functions. This
corresponds to summation over the complete set of states and does
include the continuum spectrum. Again, the result is in very good
agreement with other {\it ab initio} calculations (\cite{Lee} and
the present work).

Recent calculations by Beloy {\em et al}~\cite{Beloy} applied a mixed
approach, with extensive use of experimental data for lower cesium states
and {\em ab initio} summation over higher states including continuum.
The result is in good agreement with fully {\em ab initio} calculations.

In contrast, analysis performed in \cite{Feitchner,Micalizio,Ulzega}
is limited to discrete spectrum. Adding $-0.34 \times
10^{-1}$Hz/(V/m)$^2$ (which is total {\it tail} contribution from
all three terms (\ref{nu1}), (\ref{nu2}) and (\ref{norm}) found in
our calculation) to the results of Feitchner {\it
et~al}~\cite{Feitchner} and Micalizio {\it et~al}~\cite{Micalizio}
brings them to excellent agreement with {\it ab initio}
calculations. The same modification of the result by Ulzega {\it et
al}~\cite{Ulzega} makes it a little bit too large but still closer
to other results than without the {\it tail} contribution.

\begin{table}
    \caption{Electrostatic frequency shifts for the hyperfine
    transition of  Cs ($\delta\nu_{0}$/$E^{2} \times 10^{-10}$ Hz/(V/m)$^2$) ; comparison with other calculations and measurements.}
    \begin{ruledtabular}
        \begin{tabular}{ l l l l l}
         \multicolumn{1}{c}{This} & \multicolumn{1}{c}{Other} & \multicolumn{1}{c}{Ref} &    \multicolumn{1}{c}{Measurements} & \multicolumn{1}{c}{Ref} \\
        \multicolumn{1}{c}{work} & \multicolumn{1}{c}{calculations} &        \\
        \hline

        -2.26(2)   & -1.9(2)  & \cite{Feitchner} & -2.29(7)  & \cite{Haun}  \\
                & -2.2302  & \cite{Lee}       & -2.25(5)  & \cite{Mowat} \\
                & -2.28    & \cite{Palchikov} & -2.17(26) & \cite{Bauch} \\
                & -1.97(9) & \cite{Micalizio} & -2.271(4) & \cite{Simon} \\
                & -2.06(1) & \cite{Ulzega}    & -1.89(12) & \cite{Levi} \\
                & -2.268(8)& \cite{Beloy}     & -2.03(4)  & \cite{Godone} \\

                 \end{tabular}
    \label{results}
        \end{ruledtabular}
\end{table}

We are grateful to S. Ulzega, W. Itano, and A. Derevianko for useful comments and references.

\end{document}